\documentstyle[aps,prl,preprint,graphicx]{revtex}
\newcommand{\aand}{\mbox{ and }}

\newcommand{\Hamil}{{\cal H}}
\newcommand{\ket}[1]{|{#1}\rangle}
\newcommand{\bra}[1]{\langle{#1}|}
\newcommand{\X}[2]{X^{#1}_{#2}}

\newcommand{\noneqgronline}[2]{
	(-i)\langle {\rm T}S{#1}{#2}\rangle/
		 \langle {\rm T}S \rangle}

\newcommand{\Occu}[1]{\av{{\rm T}{#1}}}

\newcommand{\av}[1]{\langle{#1}\rangle}

\newcommand{\com}[2]{[{#1},{#2}]}
\newcommand{\anticom}[2]{\{{#1},{#2}\}}
\newcommand{\ddt}{\frac{\partial}{\partial t}}

\def\ket#1{|#1\rangle}
\def\bra#1{\langle#1|}

\def\dmu#1{{\rm d}#1}

\def\leade#1{\varepsilon_{#1\sigma}}
\def\dote#1{\varepsilon_{#1}}

\def\c#1{c_{#1\sigma}}
\def\cdagger#1{c_{#1\sigma}^{\dagger}}

\def\up{\uparrow}
\def\down{\downarrow}

\def\im{{\rm Im}}

\tightenlines

\begin{document}
%\preprint{Draft for PRL,\ \today}
%\frontmatter{
\title{Many-body approach to spin-dependent transport in quantum dot systems}
\author{J. Fransson$^1$, O. Eriksson$^1$, and I. Sandalov$^{1,2}$ }
\address{$^1$ Condensed Matter Theory group, Uppsala University, Box 530, 751 21\ \ Uppsala, Sweden\\
$^2$ Kirensky Institute of Physics, RAS, 660036 Krasnoyarsk, Russian Federation}
%\date{\today}
\maketitle

\begin{abstract}
By means of a diagram technique for Hubbard operators we show the
existence of a spin-dependent renormalization of the localized levels
in an interacting region, \emph{e.g.} quantum dot, modeled by the
Anderson Hamiltonian with two conduction bands. It is shown that the
renormalization of the levels with a given spin direction is due to
kinematic interactions with the conduction sub-bands of the opposite
spin. The consequence of this dressing of the localized levels is a
drastically decreased tunneling current for ferromagnetically ordered
leads compared to that of paramagnetically ordered leads. Furthermore,
the studied system shows a spin-dependent resonant tunneling behaviour
for ferromagnetically ordered leads.
\vspace{2mm}
\vspace{2mm}\\
\end{abstract}
%}

Spin-dependent tunneling \cite{johnson1998} and tunneling magneto
resistance (TMR) \cite{julliere1975,moodera1995} have recently been
studied extensively. Concerning spin-dependent tunneling through a
quantum dot (QD), or similar interacting regions, the main focus has
been to investigate the effects of a magnetic field applied over the
interacting region
\cite{meir1991,pustilnik2000,giuliano2000,wiel2000}. The opportunity
of changing the magnetic properties of the leads, leading in and out
of the QD, by an external magnetic field or by spin injection and
thereby altering the output current, has so far been a peripheral
topic. There are theoretical reports of spin filters and spin memories
\cite{recher2000} in which the spin polarized current is controlled by
the Zeeman splitting of the localized levels in the QD. Another
suggestion is a three-terminal system in which two of the leads are in
anti-ferromagnetic order \cite{brataas2000}. The source-drain current
is manipulated by the magnetization direction in the third
terminal. However, these studies are formulated in terms of
single-electron properties and, in addition, they cannot be directly
transformed into a time-dependent situation. To our knowledge, there
is no theoretical report of inducing a large spin-polarization in the
QD by simply spin-polarizing the conduction band.

In this Letter we demonstrate, from a many-body approach, that there
is a large spin-dependent renormalization of the levels in an
interacting region, \emph{e.g.} a QD, due to the magnetic properties
of the leads, which could be used in magnetic sensor applications. By
shifting the magnetic ordering in the leads, from paramagnetic to
ferromagnetic, the levels in the interacting region experiences a spin
split due to kinematic interactions with the conduction bands. In
fact, as we will show, the conduction electrons with the spin
projection $\sigma$ interact kinematically with the localized level
of the opposite spin $\bar{\sigma}$. This effect, in turn, causes a
drastic increase (up to 150$\%$) of the tunneling current through the
interacting region when the conduction bands are changed from a
ferromagnetic to paramagnetic ordering. Having this result at hand, we
suggest a single electron device that is sensitive to the magnetic
ordering in the contacts and is responding with an altered output
current.

To be specific, we are interested in an interacting region with a
single level, that is taking part in the conduction, in the presence
of a large Coulomb repulsion $U$.  Such a system corresponds to the
experimental reality of a QD in the Coulomb blockade state at low
temperatures $k_BT<U$, where $k_B$ is the Boltzmann contant, and small
voltages \cite{CoulombBlockade}.  The interacting region is coupled,
via tunneling (mixing) interactions $v_{k\sigma}$, to two contact
leads characterized by free electrons and the chemical potentials
$\mu_L\aand\mu_R$ for the left (L) and the right (R) leads,
respectively. A voltage applied over the system giving rise to a
difference $\mu_L-\mu_R\neq0$ results in a charge current from the
higher to the lower chemical potential. The system can be realized
with the degenerate Anderson Hamiltonian
\cite{anderson1961}, with two conduction bands, in which the localized
states in the interacting region are described by
$\Hamil_D=\sum_{p}E_{p}\X{pp}{}$, where the Hubbard operator
$\X{pp'}{}\equiv\ket{p}\bra{p'}$ represents the transition from the
state $\ket{p'}$ to $\ket{p}$ \cite{hubbard1963}. The summation is
taken over the state labels $p\in\{0,\up,\down\}$ ($p=0$ corresponds
to the local vacuum whereas the doubly occupied state $\ket{\up\down}$
is excluded because of the large $U$). A conduction electron with the
energy $\leade{k}$ in the lead $\alpha=L,R$ is created (annihilated)
by $\cdagger{k}\ (\c{k})$, $k\sigma\in\alpha$. The Hamiltonian of the
system can be written as
\begin{equation}
\Hamil=\sum_{k\sigma\in L,R}\leade{k}\cdagger{k}\c{k}
	+\sum_{p}E_{p}\X{pp}{}
	+\sum_{k\sigma}(v_{k\sigma}\cdagger{k}\X{0\sigma}{}+H.c.).
\label{eq-Anderson}
\end{equation}
The dynamics of the operator $\X{0\sigma}{}$ is given by the
Heisenberg equation of motion,
\[ i\ddt\X{0\sigma}{}=\com{\X{0\sigma}{}}{\Hamil}=
	\Delta_{\sigma0}^0\X{0\sigma}{} 
	+\sum_k[v_{k\sigma}^*(\X{00}{}+\X{\sigma\sigma}{})\c{k}
	+v_{k\bar{\sigma}}^*\X{\bar{\sigma}\sigma}{}c_{k\bar{\sigma}}]
\]
(note that $\sigma$ and $\bar{\sigma}$ denote opposite spin
projections). It is the last term in this expression that gives the
\emph{mixing induced} spin-dependent dressing of the localized
level. For a clarification of this fact, let us consider the
difference of the diagram expansions for standard Fermion operators
and Hubbard operators. When one is dealing with the expectation value
of $N$ operators, $\pi_N=\Occu{\prod_{i=1}^N\eta_i}$, Wick's
decoupling of two operators $\eta_i,\eta_j$ results in the
anticommutator $\anticom{\eta_i}{\eta_j}$. In the case of standard
Fermion operators this anticommutator is a scalar, $c_{ij}$, and the
number of operators in the expectation value is, therefore, decreased
by two, $\pi_N=\sum c_{ij}F_{ij}\pi^{ij}_{N-2}$, where
$F_{ij}=\Occu{\eta_i\eta_j}$ is a Fermion propagator. Now, in the case
of Hubbard operators the anticommutator becomes yet again an operator,
$P_{ij}$. The number of operators in the expectation $\pi_N$ value is,
thereby, only reduced by one, $\pi_N=\sum D_{ij}\pi^{ij}_{N-1}$, where
$D_{ij}=\Occu{\eta_i\eta_j}$ is a propagator of Hubbard operators,
implying that one has to make a decoupling also with $P_{ij}$, since
it remains in the expectation value $\pi_{N-1}$. The terms in the
perturbation expansion coming from the decouplings with $P_{ij}$ give
rise to the \emph{kinematic} interactions and are characteristic for
strongly correlated electron systems. In our particular case the
decoupling in the first step gives
$\anticom{\X{0\sigma}{}}{\sum_s\X{s0}{}c_{ks}}=(\X{00}{}+\X{\sigma\sigma}{})\c{k}+\X{\bar{\sigma}\sigma}{}c_{k\bar{\sigma}}$,
whereas the kinematic interactions are, in the second step, generated
by the commutator
$\com{\X{0\sigma}{}}{(\X{00}{}+\X{\bar{\sigma}\bar{\sigma}}{})c_{k\bar{\sigma}}+\X{\sigma\bar{\sigma}}{}\c{k}}=\X{0\bar{\sigma}}{}\c{k}$.
This effect is clearly seen to arise solely due to correlations.

The density of electron states (DOS) for each spin projection $\sigma$
in the interacting region is given by $\rho_{\sigma}(\omega)=-1/\pi\im
G_{\sigma\sigma}^r(\omega)$, where the Green function (GF)
$G_{\sigma\sigma'}(t,t')\equiv
\noneqgronline{\X{0\sigma}{}(t)}{\X{\sigma'0}{}(t')}$, with
$S=exp(-i\int_0^{-i\beta}\Hamil'(t)\dmu{t})$. The Hamiltonian
\[ \Hamil'(t)=
	U_0(t)\X{00}{}
	+\sum_{\sigma}[U_{\sigma}(t)\X{\sigma\sigma}{}
	+U_{\sigma\bar{\sigma}}(t)\X{\sigma\bar{\sigma}}{}]
\]
is a time-dependent disturbance to the system, by which a perturbation
expansion of $G_{\sigma\sigma'}$ is generated through functional
differentiation with respect to the fields $U_{\xi}(t)$
\cite{sandalov1995}. The equation of motion for the GF of the
interacting region is
\begin{eqnarray}
\left(i\ddt-\Delta^0_{\sigma0}-\Delta U_{\sigma0}(t)\right)	
G_{\sigma\sigma'}(t,t')&&
	-U_{\sigma\bar{\sigma}}(t)G_{\bar{\sigma}\sigma'}(t,t')
	=\delta(t-t')P_{\sigma\sigma'}(t)
\nonumber\\
&&	+[P_{\sigma\sigma}(t^+)+R_{\sigma\sigma}(t^+)]
		\int_0^{-i\beta}
		V_{\sigma}(t,t'')G_{\sigma\sigma'}(t'',t')\dmu{t''}
\nonumber\\
&&	+[P_{\sigma\bar{\sigma}}(t^+)+R_{\sigma\bar{\sigma}}(t^+)]
		\int_0^{-i\beta}
		V_{\bar{\sigma}}(t,t'')G_{\bar{\sigma}\sigma'}(t'',t')\dmu{t''}.
\label{eq-QDGF}
\end{eqnarray}
Here $\Delta^0_{\sigma0}=E_{\sigma}-E_0$ is the bare transition
energy, $\Delta U_{\sigma0}(t)=U_{\sigma}(t)-U_0(t)$ and $R_{\sigma\sigma'}(t)\equiv
\delta_{\sigma\sigma'}i\delta/\delta U_{0}(t)+i\delta/\delta
U_{\sigma'\sigma}(t)$. The expectation value
$P_{\sigma\sigma'}=\av{{\rm
T}\anticom{\X{0\sigma}{}}{\X{\sigma'0}{}}}/\av{{\rm
T}S}=\delta_{\sigma\sigma'}N_0+N_{\sigma'\sigma}$ is the sum of the
population numbers $N_0$ and $N_{\sigma'\sigma}$ corresponding to the
transitions $[00]$ and $[\sigma'\sigma]$, respectively. Physical
quantities are obtained as $U_{\xi}(t)\rightarrow0$. In this limit,
all expectation values which do not conserve the longitudinal
component of the total spin vanish, although their functional
derivatives may not. The propagator $V_{\sigma}(t,t')=\sum_{k\in
L,R}|v_{k\sigma}|^2g_{k\sigma}(t,t')$, where $g_{k\sigma}$ is the GF
of free electrons in the lead $\alpha$.

\begin{figure}[tb]
\begin{center}
\includegraphics[width=7.5cm]{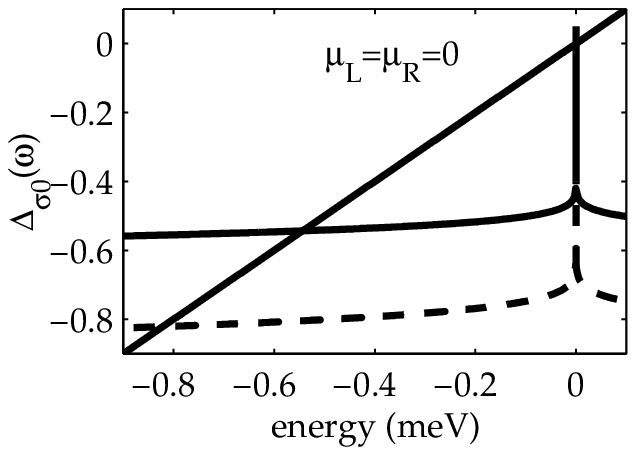}
\end{center}
\caption{The graphical solution of the renormalization Eq. (\ref{eq-renormalization}) for the spin $\up$ (solid) and spin $\down$ (dashed) level in the interacting region. The bare transition energy $\Delta^0_{\sigma0}=-0.1$ relative to the chemical potential $\mu_{\alpha}=0$, the coupling strength $\Gamma_{\sigma}^{\alpha}=2\pi|v_{\sigma}|^2\rho_{\sigma}^{\alpha}=0.5$ and the temperature $kT=0.175$. The conduction band density of states $\rho_{\sigma}^{\alpha}=1/2W=1/100$. The spin polarization in the conduction band is given by the lower cut offs $-W_{\up}^{\alpha}=-50, -W_{\down}^{\alpha}=-12.5$ (units: meV).}
\label{fig-transitionenergies}
\end{figure}

\begin{figure}[tb]
\begin{center}
\includegraphics[width=7.5cm]{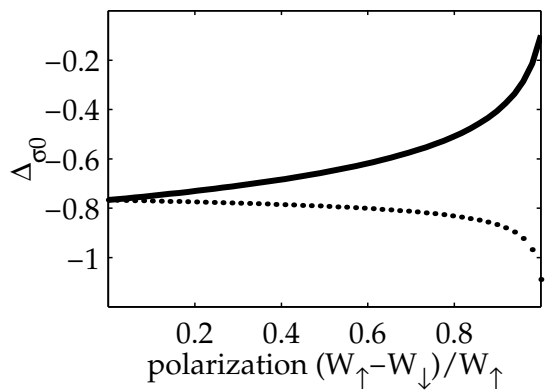}
\end{center}
\caption{The dressed transition energies $\Delta_{\up0}$ (solid) and $\Delta_{\down0}$ (dashed) as a function of the spin polarization $(W_{\up}^{\alpha}-W_{\down}^{\alpha})/W_{\up}^{\alpha}$ in the conduction bands. The polarization in the two leads are equal.}
\label{fig-spindeprenorm}
\end{figure}

We look for a GF of the form \cite{sandalov1995}
$G_{\sigma\sigma'}(t,t')=D_{\sigma\sigma'}(t,t')P_{\sigma\sigma'}(t')$,
where the locator $D_{\sigma\sigma'}$ provides the essential physical
information when all the $P_{\xi}$ are approximated by constants. A
more detailed study \cite{franssonKondo} shows that taking into
account effects of $\delta P_{\xi}(t)/\delta U_{\xi'}(t')$ only
marginally modify our results. If we neglect all functional
derivatives in Eq. (\ref{eq-QDGF}) the Hubbard-I approximation
\cite{sandalov1995} is recovered. By also calculating the first
functional derivative of the GF, for which the only non-vanishing
contribution is
\[ R_{\sigma\bar{\sigma}}(t^+)G_{\bar{\sigma}\sigma}(t'',t')=
	-iD_{\bar{\sigma}\bar{\sigma}}(t'',t_1)
	\Gamma_0(t_1,t_2;t^+)
	G_{\sigma\sigma}(t_2,t'),
\]
we find the first order equation in the tunneling interaction $V$.
Here, we have defined the zero vertex $\Gamma_0(t_1,t_2;t^+)=\delta
d_{\bar{\sigma}\sigma}^{-1}(t_1,t_2)/\delta
U_{\bar{\sigma}\sigma}(t^+)=-\delta(t_1-t_2)\delta(t_2-t^+)$
\cite{sandalov1995}, where $d$ is the locator of the interacting
region for vanishing tunneling interactions with the leads. The Dyson
equation for the locator $D_{\sigma\sigma}$ generated by the zero
vertex, the \emph{loop correction}, is graphically given by
\begin{center}
\includegraphics[width=7.5cm]{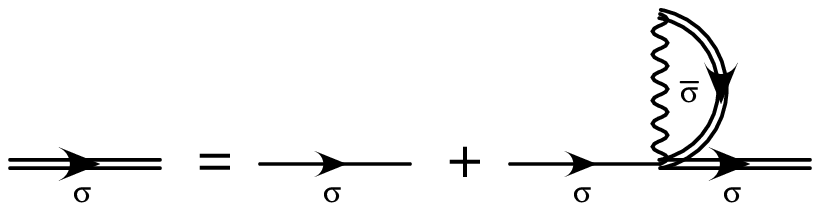}
\end{center}
where the single and double straight lines symbolize the locators $d$
and $D$, respectively. The wiggly line denotes the effective
interaction $V$. We draw attention to the fact that the localized
level with the spin projection $\sigma$ interacts kinematically only
with the conduction electrons of the opposite spin $\bar{\sigma}$ and
it is this effect that gives the possibility of a large magneto
resistance (MR). The renormalized transition energy $\Delta_{\sigma0}$
is given by the equation
\begin{equation}
\Delta_{\sigma0}-\Delta_{\sigma0}^0=
	-\sum_{k\in L,R}|v_{k\bar{\sigma}}|^2
	\frac{f(\Delta_{\bar{\sigma}0})-f(\dote{k\bar{\sigma}})}
	{\Delta_{\bar{\sigma}0}-\dote{k\bar{\sigma}}},
\label{eq-renormalization}
\end{equation}
where $f(\omega)$ is the Fermi-Dirac distribution function. Note that
results similar to Eq. (\ref{eq-renormalization}) have been obtained
earlier by other methods for different models \emph{in equilibrium}
\cite{examples}. However, in none of these earlier studies the
explicit spin-dependence on the conduction electrons of the opposite
spin projection, which is present in Eq. (\ref{eq-renormalization}),
was found. For constant mixing $v$ and conduction band density of
states $\rho_{\sigma}^{\alpha}$ the shift, given by
Eq. (\ref{eq-renormalization}), clearly has a logarithmic divergence
at the chemical potential of the lead $\alpha$. Obviously then, for
certain choices of parameters Eq. (\ref{eq-renormalization}) has more
than one solution, as illustrated in
Fig. \ref{fig-transitionenergies}. Since the renormalization of
$\Delta_{\up0}$ (solid line) depends on the dressed transition energy
$\Delta_{\down0}$ (dashed line), there may be several divergences
around the chemical potential. All such solutions correspond to
possible excitations of the QD. However, the interesting solution for
each spin is that with the lowest energy. In
Fig. \ref{fig-spindeprenorm} the dressed transition energies
$\Delta_{\up0}$ (solid line) and $\Delta_{\down0}$ (dashed line) are
plotted as a function of the spin polarization in the leads, defined
as the fraction $(W_{\up}^{\alpha}-W_{\down}^{\alpha})/W$, where
$W_{\sigma}^{\alpha}$ is the high energy cut off for the constant
conduction band density of states in the lead $\alpha$ and $W$ is half
the bandwidth of the conduction band. Throughout this Letter we
consider only the case when the polarizations in the two leads are the
same. For non-polarized leads, the localized spin $\up$- and spin
$\down$-levels collapse into a two-fold degenerate level. As the spin
polarization in the leads becomes non-zero, the dressed transition
energies for the two levels become distinct and as the polarization
increases, the renormalization of the $\up$-level decreases. In the
limit of completely spin polarized conduction bands, the
renormalization vanishes and
$\Delta_{\up0}\rightarrow\Delta_{\up0}^0$.

The splitting of the localized levels in the interacting region due to
the magnetic properties in the conduction bands directly influences
the tunneling current. Of particular interest is a comparison of the
cases when the two leads are in either paramagnetic or ferromagnetic
order, since these are the relevant states in magnetic sensors. Below
we show that the different magnetic phases of the leads imply a severe
change in the magnitude of the tunneling current through the system.

\begin{figure}[tb]
\begin{center}
\includegraphics[width=7.5cm]{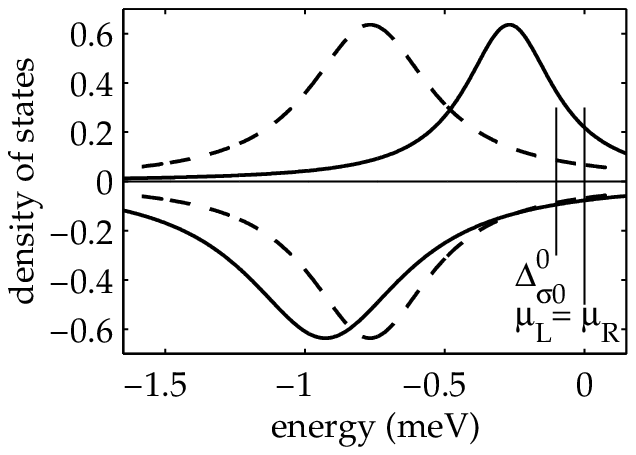}
\end{center}
\caption{The equilibrium DOS of the interacting region for paramagnetically (dashed) and ferromagnetically (solid) ordered leads. In the ferromagnetically ordered case there is an amount of $3.5\%$ of the minority spin in the leads.}
\label{fig-spinpoldos}
\end{figure}

In the stationary regime, the tunneling current through the
interacting region, symmetrically coupled to the leads, is given by
(for a detailed discussion see Refs. \cite{hershfield1991,meir1992}),
\[ J=\frac{e}{\hbar}\sum_{\sigma}\int_{-W_{\sigma}}^{-W_{\sigma}+2W}
		\Gamma_{\sigma}
	\Bigl(f_L(\dote{})-f_R(\dote{})\Bigr)
	\rho_{\sigma}(\dote{})\dmu{\dote{}},
\]
where
$\Gamma_{\sigma}=\Gamma_{\sigma}^L\Gamma_{\sigma}^R/(\Gamma_{\sigma}^L+\Gamma_{\sigma}^R)$,
$\Gamma_{\sigma}^{\alpha}=\Gamma_{\sigma}^{\alpha}(\omega)|_{\mu_{\alpha}}=2\pi|v_{\sigma}|^2\rho_{\sigma}^{\alpha}$
and $f_{\alpha}(\dote{})=f(\dote{}-\mu_{\alpha})$, which has proven
successful in the regime we consider \cite{sivan1996}. In the given
approximation the retarded GF is
\[ G_{\sigma\sigma}^r(\omega)=
	\frac{P_{\sigma\sigma}}
	{\omega-\Delta_{\sigma0}+
		i(\Gamma_{\sigma}^L+\Gamma_{\sigma}^R)
					P_{\sigma\sigma}/2},
\]
with $\Delta_{\sigma0}$ given by Eq. (\ref{eq-renormalization}) and
$P_{\sigma\sigma}=N_0+N_{\sigma}$, where $N_{\sigma}=-1/\pi\int
f(\omega)\im G_{\sigma\sigma}^r(\omega)\dmu{\omega}$ and
$N_0+N_{\up}+N_{\down}=1$. The corresponding DOS is shown in
Fig. \ref{fig-spinpoldos} where the spin $\up$ and spin $\down$ are
plotted on the positive and negative vertical axes, respectively. When
the leads are in a paramagnetic state (dashed lines), the dressed
transition energies coincide having an equal probability. For
ferromagnetically ordered leads (solid lines) the
$\Delta_{\down0}$-transition becomes more likely
($P_{\down\down}=0.69$) than the $\Delta_{\up0}$-transition
($P_{\up\up}=0.39$). At the same time the transition to the
$\down$-level retains the strong influence from the spin $\up$
conduction electrons and therefore remains as large, or larger, as in
the paramagnetic configuration.

\begin{figure}[tb]
\begin{center}
\includegraphics[width=7.5cm]{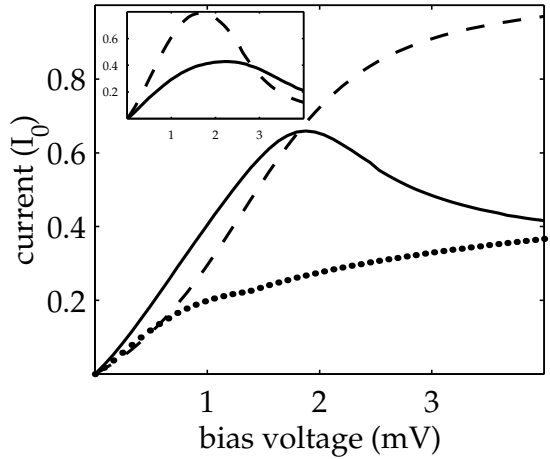}
\end{center}
\caption{The tunneling current through the interacting region for paramagnetically (dashed) and ferromagnetically ordered leads. For the ferromagnetic ordering the minority spin percentage is 3.5 (solid) and 0 (dotted). The inset shows the situation where the conduction band width is $2W=2.8$ meV. $I_0=I^{\up}_0+I^{\down}_0,\ I^{\sigma}_0=e\Gamma_{\sigma}/\hbar$.}
\label{fig-spinpolcurrent}
\end{figure}

As for the tunneling current through the system, there is a huge
discrepancy in the current for a range of voltages in the two cases,
displayed in Fig. \ref{fig-spinpolcurrent}. In
Fig. \ref{fig-spinpolcurrent} the current-voltage characteristics are
shown for three cases in which the leads are in paramagnetic (dashed
line) and ferromagnetic ordering with a minority spin percentage of
3.5 (solid line) and 0 (dotted). For sufficiently small voltages the
magnitude of the current is larger for ferromagnetic than for
paramagnetic leads. As the voltage increases, though, the current
becomes larger in the paramagnetic case and for certain voltages the
change in the MR $|R_{fm}-R_{pm}|/R_{pm}$ can be as large as 150$\%$,
a large number in view of existing experimental devices
\cite{garcia1999}.

The solid line in Fig. \ref{fig-spinpolcurrent}, describing a
spin-dependent resonant tunneling behaviour, represents a situation
where the spin polarization in the leads is such that there is only a
tiny fraction of the minority spin state present. As the bias voltage
is increased the bottom of that sub-band eventually separates from the
corresponding level in the interacting region which gives a decreasing
contribution from the minority spin to the tunnel current. This
results in a tunnel current through the system that equals the current
of the majority spin state only. The inset of
Fig. \ref{fig-spinpolcurrent} illustrates a non-intuitive and extreme
case of this situation with very narrow conduction bands. Then, for a
certain voltage range the current is \emph{decreased} as the
conduction bands are shifted from ferromagnetic to paramagnetic
configuration, giving an up to 45$\%$ \emph{inverse} MR.

In conclusion, using the Anderson model we predict that the localized
level with a given spin state in a QD is strongly renormalized, via
kinematic interactions, by the conduction band of the opposite spin
state. For ferromagnetic leads the levels in the QD experience a spin
split which results in a spin-dependent tunnel current through the
system. We observe a change in the MR by up to 150$\%$ as the magnetic
configuration in the leads are changed from ferromagnetic to
paramagnetic, suggesting that our findings can be used in devicing
magnetic sensors. The effect is non-trivial which is shown by the
possibility of an, up to 45$\%$, inverse MR.

Support from the G\"oran Gustafsson foundation, the Swedish National
Science Foundation (NFR and TRF) and the Swedish Foundation for
Strategic Research (SSF) is acknowledged.

\newpage
\noindent

\noindent

\noindent

\noindent

\end{document}